\documentclass[12pt]{article}
\usepackage{amssymb}

\input{tcilatex}

\begin{document}

\bigskip\ 

\bigskip\ 

\begin{center}
\textbf{MATROIDS AND P-BRANES}

\textbf{\ }

\textbf{\ }

\smallskip\ 

J. A. Nieto\footnote{%
nieto@uas.uasnet.mx}

\smallskip

\textit{Departamento de Investigaci\'{o}n en F\'{\i}sica de la Universidad
de Sonora,}

\textit{83190, Hermosillo Sonora, M\'{e}xico}

\textit{and}

\textit{Facultad de Ciencias F\'{\i}sico-Matem\'{a}ticas de la Universidad
Aut\'{o}noma}

\textit{de Sinaloa, 80010, Culiac\'{a}n Sinaloa, M\'{e}xico}\footnote{%
permanent address}

\bigskip\ 

\bigskip\ 

\textbf{Abstract}
\end{center}

A link between matroid theory and $p$-branes is discussed. The Schild type
action for $p$-branes and matroid bundle notion provide the two central
structures for such a link. We use such a connection to bring the duality
concept in matroid theory to $p$-branes physics. Our analysis may be of
particular interest in M-theory and in matroid bundle theory.

\bigskip\ 

\bigskip\ 

\bigskip\ 

\bigskip\ 

\bigskip\ 


Pacs numbers: 04.60.-m, 04.65.+e, 11.15.-q, 11.30.Ly

February, 2004

\newpage \bigskip

The matroid bundle mathematical structure [1]-[3] emerged as a natural
extension of oriented matroid theory [4]. Part of the mathematical
motivation for such a structure arose when Gelfand and MacPherson [5]
discovered a connection between matroid bundle and Pontrjagin classes.
Physically, the matroid bundle concept had led to the proposal of a new
gravitational theory called gravitoid theory [6]. Moreover, it had been
shown [7]-[9] that supergravity $D=11$, Chern-Simons theory and string
theory are closely related to matroid bundle.

Here, we are interested in discussing the possibility of linking matroids
and $p$-branes via Schild type action [10] (see Ref. [11] and also Refs.
[12]-[15]) for $p$-branes and matroid bundle notion. Our analysis may be of
particular interest in M-theory [16]-[18] and in matroid bundle theory
itself.

Consider a $p$-brane moving in a $d+1$-dimensional Minkowski space-time. We
describe the evolution of such a system by the $d+1$-scalar field
coordinates $x^{\mu }(\xi ^{a}),$ where $\mu =0,1,...,d$, which are
functions of the arbitrary parameters $\xi ^{a}$, with $a=0,1,...,p.$

The Dirac-Nambu-Goto type action for $p$-branes is

\begin{equation}
S_{p}^{(1)}=-T_{p}\int d^{p+1}\xi \sqrt{-h},  \label{1}
\end{equation}
where $h\equiv \det (h_{ab}),$ with

\begin{equation}
h_{ab}=\partial _{a}x^{\mu }\partial _{b}x^{\nu }\eta _{\mu \nu },  \label{2}
\end{equation}%
and $T_{p}$ is a fundamental constant measuring the inertia of the $p$%
-brane. Here,

\begin{equation}
\eta _{\mu \nu }=diag(-1,1,...,1)  \label{3}
\end{equation}
is the Minkowski metric.

Let us write $h$ in the form

\[
h=\frac{1}{(p+1)!}\sigma ^{\mu _{1}...\mu _{p+1}}\sigma _{\mu _{1}...\mu
_{p+1}}, 
\]
where

\begin{equation}
\sigma ^{\mu _{1}...\mu _{p+1}}=\varepsilon ^{a_{1}...a_{p+1}}v_{a_{1}}^{\mu
_{1}}(\xi )...v_{a_{p+1}}^{\mu _{p+1}}(\xi ).  \label{4}
\end{equation}
Here, $\varepsilon ^{a_{1}...a_{p+1}}$ is the totally antisymmetric tensor
and

\begin{equation}
v_{a}^{\mu }(\xi )=\partial _{a}x^{\mu }(\xi ).  \label{5}
\end{equation}

It turns out that the action (1) is equivalent to

\begin{equation}
S_{p}^{(2)}=\int d^{p+1}\xi (\sigma ^{\mu _{1}...\mu _{p+1}}p_{\mu
_{1}...\mu _{p+1}}-\frac{\gamma }{2}(p^{\mu _{1}...\mu _{p+1}}p_{\mu
_{1}...\mu _{p+1}}+T_{p}^{2})),  \label{6}
\end{equation}%
where $\gamma $ is a lagrange multiplier and the quantity $p_{\mu _{1}...\mu
_{p+1}}$ can be understood as the linear momentum associated to $\sigma
^{\mu _{1}...\mu _{p+1}}.$ Varying (6) with respect to $p_{\mu _{1}...\mu
_{p+1}}$ it allows to eliminate $p_{\mu _{1}...\mu _{p+1}}$. We get

\begin{equation}
S_{p}^{(3)}=\frac{1}{2}\int d^{p+1}\xi (\gamma ^{-1}\sigma ^{\mu _{1}...\mu
_{p+1}}\sigma _{\mu _{1}...\mu _{p+1}}-\gamma T_{p}^{2}).  \label{7}
\end{equation}%
By eliminating $\gamma $ from (7) one recovers the action (1). The
importance of (6) or (7) is that it now makes sense to set $T_{p}=0.$ In
this case (7) is reduced to the Schild type null $p$-brane action [10]-[11].

Here, we are interested in relating (7) to matroid bundle theory. For this
purpose it is convenient to recall the definition of an oriented matroid.

An oriented matroid $\mathcal{M}$ is a pair $(S,\chi ),$ where $S$ is a
non-empty finite set and $\chi $ (called chirotope) is a mapping $%
S^{r}\rightarrow \{-1,0,1\}$, with $r$ the rank on $S$, satisfying the
following properties.

$(\chi i)\chi $ is not identically zero,

$(\chi ii)\chi $ is alternating,

$(\chi iii)$ for all $x_{1},x_{2},...,x_{r},y_{1},y_{2},...,y_{r}\in S$ such
that

\begin{equation}
\chi (x_{1},x_{2},...,x_{r})\chi (y_{1},y_{2},...,y_{r})\neq 0,  \label{8}
\end{equation}
there exists an $i\in \{1,2,...,r\}$ such that

\begin{equation}
\chi (y_{i},x_{2},...,x_{r})\chi
(y_{1},y_{2},...,y_{i-1},x_{1},y_{i+1,}...,y_{r})=\chi
(x_{1},x_{2},...,x_{r})\chi (y_{1},y_{2},...,y_{r}).  \label{9}
\end{equation}

For a vector configuration the chirotope $\chi $ can be identified as

\begin{equation}
\chi (\mu _{1},...,\mu _{r})\equiv sign\det (b^{\mu _{1}},...,b^{\mu
_{r}})\in \{-1,0,1\},  \label{10}
\end{equation}
for all $b^{\mu _{1}},...,b^{\mu _{r}}\in R^{r}$ and for all $\mu
_{1},...,\mu _{r}\in S$. In this case (10) becomes connected with the
Grassmann-Plucker relation (see Ref.. [4], section 3.5).

It can be proved that the definition of the underlying matroid $M$ of $%
\mathcal{M}$ follows from the chirotope definition for oriented matroids. In
fact, from the chirotope definition it follows that if $\mathcal{B}$ is the
set of $r$-subsets of $S$ such that

\begin{equation}
\chi (x_{1},x_{2},...,x_{r})\neq 0,  \label{11}
\end{equation}%
for some ordering of $(x_{1},x_{2},...,x_{r})$ of $\mathcal{B}$, then $%
\mathcal{B}$ is the set of bases of the matroid $M$. Formally, the
definition of $M$ in terms of the bases is as follows (see Ref. [19]):

A matroid $M$ is a pair $(S,\mathcal{B})$, where $S$ is a non-empty finite
set and $\mathcal{B}$ is a non-empty collection of subsets of $S$ (called
bases) satisfying the following properties:

$(\mathcal{B}$ $\mathit{i)}$\textit{\ }no base properly contains another
base;

$(\mathcal{B}$ $\mathit{ii)}$ if $B_{1}$ and $B_{2}$ are bases and if $b$ is
any element of $B_{1},$ then there is an element $g$ of $B_{2}$ with the
property that $(B_{1}-\{b\})\cup \{g\}$ is also a base.

Let us write (10) in the form $\chi (\mu _{1},...,\mu _{r})\equiv sign\Sigma
^{\mu _{1}...\mu _{r}},$ where 
\begin{equation}
\Sigma ^{\mu _{1}...\mu _{r}}\equiv \varepsilon
^{a_{1}...a_{r}}b_{a_{1}}^{\mu _{1}}...b_{a_{r}}^{\mu _{r}}.  \label{12}
\end{equation}%
Here, the indices $a_{1},...,a_{r}$ run from $1$ to $r$. Comparing (4) and
(12) we observe the great similarity between the two formulae. The main
difference comes from the fact that while $v_{a}^{\mu }(\xi )$ is a local
object, $b_{a}^{\mu }$ is not. Therefore, our task is to understand the
transition from $b_{a}^{\mu }$ to $v_{a}^{\mu }(\xi )$.

Let $\wedge _{r}R^{n}$ denote the $(_{r}^{n})$-dimensional real vector space
of alternating $r$-forms on $R^{n}$. An element $\mathbf{\Sigma }$ in $%
\wedge _{r}R^{n}$ is said to be decomposable if

\begin{equation}
\mathbf{\Sigma }=\mathbf{b}_{1}\wedge \mathbf{b}_{2}\wedge ...\wedge .%
\mathbf{b}_{r},  \label{13}
\end{equation}
for some $\mathbf{b}_{1},\mathbf{b}_{2},...,.\mathbf{b}_{r}\in R^{n}$. It is
not difficult to see that (13) can be written as

\begin{equation}
\mathbf{\Sigma }=\frac{1}{r!}\Sigma ^{\mu _{1}...\mu _{r}}\omega _{\mu
_{1}}\wedge \omega _{\mu _{2}}\wedge ...\wedge \omega _{\mu _{r}},
\label{14}
\end{equation}%
where $\omega _{\mu _{1}},\omega _{\mu _{2}},...,\omega _{\mu _{r}}$ are one
form bases in $R^{n}$ and $\Sigma ^{\mu _{1}...\mu _{r}}$ is given in (12).
This shows that $\Sigma ^{\mu _{1}...\mu _{r}}$ can be identified with an
alternating decomposable $r$-form. \ It is known that the projective variety
of decomposable forms is isomorphic to the Grassmann variety of $r$%
-dimensional linear subspaces in $R^{n}$. In turn, the Grassmann variety is
the classifying space for vector bundle structures. These simple
observations may motivate one to look for a link between matroid theory and
vector bundle formalism.

Fortunately, the mathematicians have already developed the matroid bundle
concept [1]-[4]. The central idea in matroid bundles, introduced by
MacPherson [1], is to replace tangent spaces in a differential manifold by
oriented matroids. Specifically, one starts with a simplicial complex $X$
associated to a differential manifold $B$ by the smooth triangulation $\eta
:\shortparallel X\shortparallel \rightarrow B$. One considers the linear map 
$f_{\xi }:\shortparallel star\Delta \shortparallel \rightarrow U\subset
T_{\eta (\xi )}$ such that $f_{\xi }(\xi )=0$, where $\shortparallel \Delta
\shortparallel $ is the minimal simplex of $\shortparallel X\shortparallel $
containing $\xi \in X.$ Then, $f_{\xi }\shortparallel (star\Delta
)^{0}\shortparallel $, where $(star\Delta )^{0}$ are the $0$-simplices of $%
star\Delta ,$ is a configuration of vectors in $T_{\eta (\xi )}$ defining an
oriented matroid $\mathcal{M}(\xi )$. (For a more precise definition of
matroid bundle, see [1] and [3].)

Suppose we identify the differential manifold $B$ with the world-volume of a 
$p$-brane. According to our previous discussion one can associate an
oriented matroid $\mathcal{M}(\xi )$ at each point $\xi $ of $X$ via the
configuration of vectors given by the map $f_{\xi }\shortparallel
(star\Delta )^{0}\shortparallel $. If we consider the oriented matroid $%
\mathcal{M}(\xi )$ in terms of $(S,\chi )$ with $\chi (\mu _{1},...,\mu
_{r})\equiv sign\Sigma ^{\mu _{1}...\mu _{r}}$ we discover that the function 
$f_{\xi }$ should induce a map

\begin{equation}
\Sigma ^{\mu _{1}...\mu _{r}}\rightarrow \sigma ^{\mu _{1}...\mu _{p+1}}(\xi
),  \label{15}
\end{equation}
where we consider that the rank $r$ of $\mathcal{M}(\xi )$ is $r=p+1$. Note
that the formula (15) means that the function $f_{\xi }$ also induces the
map $b_{a}^{\mu }$ $\rightarrow v_{a}^{\mu }(\xi ).$

Our last task is to find a mechanism to go from (4) to (5). Consider the
expression

\begin{equation}
F_{ab}^{\mu }=\partial _{a}v_{b}^{\mu }(\xi )-\partial _{b}v_{a}^{\nu }(\xi
).  \label{16}
\end{equation}%
If the object $F_{ab}^{\mu }$ vanishes, then a solution of (16) is $%
v_{a}^{\mu }(\xi )=\frac{\partial x^{\mu }}{\partial \xi ^{a}},$ where $%
x^{\mu }$ is in this context a gauge function. In this case, one says that $%
v_{a}^{\mu }(\xi )$ is a pure gauge. Of course, $F_{ab}^{\mu }$ and $%
v_{b}^{\mu }(\xi )$ can be interpreted as field strength and abelian gauge
potential, respectively. Using the Palatini formalism, the formula 
\begin{equation}
F_{ab}^{\mu }=0  \label{17}
\end{equation}%
can be imposed in the action (7) as a constraint. In two dimensions, such a
formula may be derived from the abelian Chern-Simons action

\begin{equation}
S_{CS}=\frac{k}{2\pi }\int d^{3}\xi \varepsilon ^{ijk}v_{i}^{\mu }F_{jk\mu }.
\label{18}
\end{equation}
The expressions (15) and (17) can be considered as the key bridge to link $p$%
-branes and matroid bundles. It is worth mentioning that according to the
action (7), our results also apply to null $p$-branes.

It is interesting to observe that $\sigma ^{\mu _{1}...\mu _{p+1}}(\xi )$ is
a decomposable $p+1-$form. Thus, $\sigma ^{\mu _{1}...\mu _{p+1}}(\xi )$ is
connected to the Grassmann variety concept associated to a matroid bundle.
In the literature this kind of Grassmann variety is called MacPherson's
variety. It turns out that the MacPherson variety plays the same role for
matroid bundles as the ordinary Grassmann variety plays for vector bundles
(see Ref. [3] for details).

One of the attractive features that arises from the above link between
matroids and $p$-branes is that the concept of duality becomes part of the $%
p $-brane structure. The reason for this is that every oriented matroid $%
\mathcal{M}(\xi )$ has an associated unique dual oriented matroid $\mathcal{M%
}^{\ast }(\xi )$ (see Ref. [4], section 3.4) and therefore the
identification of $\sigma ^{\mu _{1}...\mu _{p+1}}(\xi )$ with the chirotope 
$\chi $ of $\mathcal{M}(\xi )$ should imply an identification of the dual of 
$\sigma ^{\mu _{1}...\mu _{p+1}}(\xi )$ with the dual chirotope $\chi ^{\ast
}.$ In order to be more specific in these observations we need to resort on
the duality concept in matroid bundle. Unfortunately, it seems that the
mathematicians have not yet considered such a concept. Nevertheless, it is
tempting to try to outline the main idea. Consider the dual of $\Sigma ^{\mu
_{1}...\mu _{p+1}}$

\begin{equation}
^{\ast }\Sigma ^{\mu _{p+2}...\mu _{d+1}}=\frac{1}{(p+1)!}\varepsilon _{\mu
_{1}...\mu _{p+1}}^{\mu _{p+2}...\mu _{d+1}}\Sigma ^{\mu _{1}...\mu _{p+1}}.
\label{19}
\end{equation}
In order to identify $^{\ast }\Sigma ^{\mu _{p+2}...\mu _{d+1}}$ with the
corresponding chirotope $\chi ^{\ast }$ we should have

\begin{equation}
^{\ast }\Sigma ^{\mu _{p+2}...\mu _{d+1}}=\varepsilon ^{\hat{a}_{p+2}...\hat{%
a}_{d+1}}b_{\hat{a}_{p+2}}^{\mu _{p+2}}...b_{\hat{a}_{d+1}}^{\mu _{d+1}},
\label{20}
\end{equation}
where the indices $\hat{a}_{p+2,}...,\hat{a}_{d+1}$ run from $p+2$ to $d+1$.
Therefore, using (12) and (20) we discover that (19) becomes

\begin{equation}
\varepsilon ^{\hat{a}_{p+2}...\hat{a}_{d+1}}b_{\hat{a}_{p+2}}^{\mu
_{p+2}}...b_{\hat{a}_{d+1}}^{\mu _{d+1}}=\frac{1}{(p+1)!}\varepsilon _{\mu
_{1}...\mu _{p+1}}^{\mu _{p+2}...\mu _{d+1}}\varepsilon
^{a_{1}...a_{p+1}}b_{a_{1}}^{\mu _{1}}...b_{a_{p+1}}^{\mu _{p+1}}.
\label{21}
\end{equation}%
An important duality property is that the vectors $b_{\hat{a}}^{\mu }$ and $%
b_{a}^{\mu }$ should satisfy the orthogonality condition (see Ref. [19])

\begin{equation}
b_{\hat{a}}^{\mu }b_{a\mu }=0.  \label{22}
\end{equation}

In order to make sense of the formula (21) at the level of matroid bundle we
need to consider the maps $b_{a}^{\mu }$ $\rightarrow v_{a}^{\mu }(\xi )$
and $b_{\hat{a}}^{\mu }$ $\rightarrow v_{\hat{a}}^{\mu }(\xi ).$ In this
case (21) becomes

\begin{equation}
\varepsilon ^{\hat{a}_{p+2}...\hat{a}_{d+1}}v_{\hat{a}_{p+2}}^{\mu
_{p+2}}(\xi )...v_{\hat{a}_{d+1}}^{\mu _{d+1}}(\xi )=\frac{1}{(p+1)!}%
\varepsilon _{\mu _{1}...\mu _{p+1}}^{\mu _{p+2}...\mu _{d+1}}\varepsilon
^{a_{1}...a_{p+1}}v_{a_{1}}^{\mu _{1}}(\xi )...v_{a_{p+1}}^{\mu _{p+1}}(\xi
).  \label{23}
\end{equation}
The next step is to connect (23) with a $p$-brane and its dual. This can be
achieved by writing

\begin{equation}
v_{a}^{\mu }(\xi )=\partial _{a}x^{\mu }(\xi )  \label{24}
\end{equation}
and

\begin{equation}
v_{\hat{a}}^{\mu }(\xi )=\partial _{\hat{a}}x^{\mu }(\xi ),  \label{25}
\end{equation}
where $x^{\mu }(\xi )$ are $d+1$-scalar fields. But this means that we
should have $\xi =(\xi ^{a},\xi ^{\hat{a}})$ instead of just $\xi =(\xi
^{a}) $. In the context of fiber bundles the coordinates $\xi ^{a}$
parametrize locally the base space $B$. Therefore we are forced to identify
the coordinates $\xi ^{\hat{a}}$ with the fiber $F$ of some bundle $E$ with
base space $B$. Fortunately, this kind of scenario is possible if we
associate $v_{a}^{\mu }(\xi )$ with the horizontal part $H_{\xi }(E)$ and $%
v_{\hat{a}}^{\mu }(\xi )$ with the vertical part $V_{\xi }(E)$ of a tangent
bundle $T_{\xi }(E)$, where $\xi $ is any point in the total space $E$. In
fact in this case we have that if $v_{a}^{\mu }(\xi )\in H_{\xi }(E)$ and $%
v_{\hat{a}}^{\mu }(\xi )\in V_{\xi }(E)$ then

\begin{equation}
v_{\hat{a}}^{\mu }v_{a\mu }=0  \label{26}
\end{equation}
as required by (22).

Another interesting aspect of the matroid-brane connection is that in
matroid theory the concept of duality may be implemented at the quantum
level for different $p$-branes. In fact, an important theorem in oriented
matroid theory assures that

\begin{equation}
(\mathcal{M}_{1}\oplus \mathcal{M}_{2})^{\ast }=\mathcal{M}_{1}^{\ast
}\oplus \mathcal{M}_{2}^{\ast },  \label{27}
\end{equation}
where $\mathcal{M}_{1}\oplus \mathcal{M}_{2}$ is the direct sum of two
oriented matroids $\mathcal{M}_{1}$ and $\mathcal{M}_{2}$. If we associate $%
p_{1}-$brane and $p_{2}-$brane to the matroids $\mathcal{M}_{1}$ and $%
\mathcal{M}_{2}$ respectively, then the corresponding partition functions

\begin{equation}
Z_{1}=\int DX\exp (S_{p_{1}}^{(3)})  \label{28}
\end{equation}%
and 
\begin{equation}
Z_{2}=\int DX\exp (S_{p_{2}}^{(3)})  \label{29}
\end{equation}%
should lead to the symmetry $Z=Z^{\ast }$ of the total partition function $%
Z=Z_{1}Z_{2}.$ Here, the actions $S_{p_{1}}^{(3)}$ and $S_{p_{2}}^{(3)}$ are
determined by (7).

Before we make some final comments, let us discuss an extension of the Hodge
duality definition (19) suggested my matroid theory. We first observe that
the completely antisymmetric object $\varepsilon _{\mu _{1}...\mu _{p+1}\mu
_{p+2}...\mu _{d+1}},$ using in (19), is in fact a chirotope associated to
the underlaying uniform matroid $U_{n,n}$. It turns out that the matroid $%
U_{n,n}$ corresponds to the ground set $S=\{1,2,...,n\}$ and bases subset $%
\mathcal{B}=\{\{1,2,...,n\}\}$, with $n=d+1.$ Therefore, there is just one
base of rank $r=n$ in $\mathcal{B}$, namely the set $\{1,2,...,n\}$ itself.
Thus, the chirotope $\chi $ associated to this base set reads as $\chi (\mu
_{1},...,\mu _{d+1})$ and using (8)-(9) one may verify that $\chi (\mu
_{1},...,\mu _{d+1})$ is in fact equal to the density $\varepsilon _{\mu
_{1}...\mu _{d+1}}.$ The question arises: from many possible chirotopes, why
is the chirotope $\varepsilon _{\mu _{1}...\mu _{d+1}}$ used to define
duality? An straightforward answer to this question it may say that because
the chirotope $\varepsilon _{\mu _{1}...\mu _{d+1}}$ has the required
properties for duality. But from the point of view chirotope theory the
object $\varepsilon _{\mu _{1}...\mu _{d+1}}$ is just a very particular
example of a chirotope. Thus, we arrive to the related question: why do not
we use other chirotopes to extend the Hodge duality concept? Let us extend
(19) in the form

\begin{equation}
^{\ddagger }\Sigma ^{\mu _{p+2}...\mu _{r}}=\frac{1}{(p+1)!}\chi _{\mu
_{1}...\mu _{p+1}}^{\mu _{p+2}...\mu _{r}}\Sigma ^{\mu _{1}...\mu _{p+1}},
\label{30}
\end{equation}%
where $\chi _{\mu _{1}...\mu _{p+1}\mu _{p+2}...\mu _{r}}\equiv \chi (\mu
_{1},..,\mu _{p+1},\mu _{p+2},...,\mu _{r})$ is a chirotope associated to
some oriented matroid of rank $r\geq p+1.$ In order to emphasize that $%
^{\ddagger }\mathbf{\Sigma }$ is a more general object than $^{\ast }\mathbf{%
\Sigma }$ let us call $^{\ddagger }\mathbf{\Sigma }$ the dualoid of $\mathbf{%
\Sigma }$. Of course, (19) is a particular case of (30), since when $r=d+1$
(30) becomes (19).

As an example of such a dualoid let us consider a $2$-brane in $d+1=11$
dimensions and some oriented matroid $\mathcal{M}$ of rank $r=6$. From (30)
we have

\begin{equation}
^{\ddagger }\Sigma ^{\alpha \beta \tau }=\frac{1}{3!}\chi _{\mu \nu \rho
}^{\alpha \beta \tau }\Sigma ^{\mu \nu \rho }.  \label{31}
\end{equation}%
This leads to the interesting result that the dualoid $^{\ddagger }\Sigma
^{\alpha \beta \tau }$ may also describe a $2$-brane in eleven dimensions.
In contrast, observe that if instead of (31) we used the traditional Hodge
transformation (19) we get that the dual of a $2$-brane is a $7-$brane.

Summarizing, in this brief work we have considered the possibility to
connect oriented matroids with $p$-branes. We have shown that makes sense to
associate the $p+1$-form $\sigma ^{\mu _{1}...\mu _{p+1}}(\xi )$ in the
Schild type action (7) with the chirotope $\chi (\mu _{1},...,\mu _{p+1})$
of an oriented matroid. It should emphasized that our procedure is not just
a technical translation from $p$-branes to matroid theory which is already
interesting, but it is a bridge that may allow to bring many important
theorems and concepts in matroid theory to $p$-brane physics. In particular,
as a proof of the importance of having established such a bridge, we have
shown that the duality concept in matroid theory can be understood as a
duality symmetry in the context of $p$-branes. The fact that this symmetry
is part of the $p+1$-form/chirotope connection of the Schild type action is
a guarantee of having classically such a symmetry in a dynamic context.
However, in a quantum context one should be always careful with classical
symmetries because of some possible anomalies, but in principle this
scenario shows a possible route to investigate the duality matroid symmetry
for a $p$-brane at the quantum level.

The question arises whether the present connection between matroids and
p-branes may be useful in M-theory itself. Of course, since our analysis
applies to any $p$-brane it must also be truth for strings which are part of
M-theory structure. But beyond this observation there is a key reason to
believe that such a connection may have more implications on M-theory. The
key idea is to consider duality as a fundamental principle in M-theory. Just
as the equivalence principle in gravity suggested to look for a mathematical
structure beyond Euclidean geometry, duality in M-theory seems to require a
mathematical structure beyond the mathematical structure usually considered
in string theory. Surprisingly, such a mathematical structure seems to be
precisely matroid theory as it has come to be evident in references [6]-[9].
The main reason is that duality plays a central role in matroid theory and
in fact, in strict sense, matroid theory may even be called a duality
theory. If one assumes that in effect matroid theory is the underlaying
mathematical structure of M-theory then one should expect new duality
properties in M-theory beyond the duality symmetries interrelating the five
known supertring theories and $p$-branes. For instance, the string/5-brane
duality in ten dimensions [20] may be considered as a particular case of the
dualoid described above. In fact, the string/5-brane arises from the field
strength $F_{\mu \nu \alpha }=\partial _{\lbrack \mu }A_{\nu \alpha ]}$
associated to the antisymmetric gauge field $A_{\nu \alpha }$ in ten
dimensions. The Hodge dual of $F_{\mu \nu \alpha }$ is

\begin{equation}
F^{\mu _{1}\mu _{2}\mu _{3}\mu _{4}\mu _{5}\mu _{6}\mu _{7}}=\frac{1}{3!}%
\varepsilon ^{\mu _{1}\mu _{2}\mu _{3}\mu _{4}\mu _{5}\mu _{6}\mu _{7}\mu
_{8}\mu _{9}\mu _{10}}F_{\mu _{8}\mu _{9}\mu _{10}}.  \label{32}
\end{equation}%
Here, $F_{\mu _{1}\mu _{2}\mu _{3}\mu _{4}\mu _{5}\mu _{6}\mu _{7}}=\partial
_{\lbrack \mu _{7}}A_{\mu _{1}\mu _{2}\mu _{3}\mu _{4}\mu _{5}\mu _{6}]}$ is
the field strength associated to the completely antisymmetric gauge field $%
A_{\mu _{1}\mu _{2}\mu _{3}\mu _{4}\mu _{6}\mu _{7}}$ which in turn implies
a 5-brane structure via the coupling

\begin{equation}
\sigma ^{\mu _{1}\mu _{2}\mu _{3}\mu _{4}\mu _{5}\mu _{6}}A_{\mu _{1}\mu
_{2}\mu _{3}\mu _{4}\mu _{5}\mu _{6}}.  \label{33}
\end{equation}%
Under our considerations the object $\sigma ^{\mu _{1}\mu _{2}\mu _{3}\mu
_{4}\mu _{5}\mu _{6}}$ can be identified with a chirotope $\chi (\mu _{1}\mu
_{2}\mu _{3}\mu _{4}\mu _{5}\mu _{6})$ and therefore new duality may arise
if instead of (32) we consider the dualoid transformation

\begin{equation}
^{\ddagger }F^{\mu _{1}\mu _{2}\mu _{3}\mu _{4}\mu _{5}\mu _{6}\mu _{7}}=%
\frac{1}{3!}\chi ^{\mu _{1}\mu _{2}\mu _{3}\mu _{4}\mu _{5}\mu _{6}\mu
_{7}\mu _{8}\mu _{9}\mu _{10}}F_{\mu _{8}\mu _{19}\mu _{10}},  \label{34}
\end{equation}%
where $\chi ^{\mu _{1}\mu _{2}\mu _{3}\mu _{4}\mu _{5}\mu _{6}\mu _{7}\mu
_{8}\mu _{9}\mu _{10}}$ may be associated to some oriented matroid of rank
ten. Here, $^{\ddagger }F_{\mu _{1}\mu _{2}\mu _{3}\mu _{4}\mu _{5}\mu
_{6}\mu _{7}}=\partial _{\lbrack \mu _{7}}A_{\mu _{1}\mu _{2}\mu _{3}\mu
_{4}\mu _{5}\mu _{6}]}^{\ddagger }$ is the dualoid field strength, where $%
A_{\mu _{1}\mu _{2}\mu _{3}\mu _{4}\mu _{5}\mu _{6}]}^{\ddagger }$ is the
corresponding completely antisymmetric gauge field.

Another, possibility of relating the matroid-brane link with M-theory is via
Matrix theory. Some years ago Yoneya [15] showed that it is possible to
construct a matrix theory of M-theory from the Schild type action for
strings. The staring point in the Yoneya's work is to consider the Poisson
bracket structure

\begin{equation}
\{x^{\mu },x^{\nu }\}=\frac{1}{\gamma }\sigma ^{\mu \nu },  \label{35}
\end{equation}%
where $\gamma $ is an auxiliary field. This identification suggests to
replace the Poisson structure by a coordinates operators

\begin{equation}
\{x^{\mu },x^{\nu }\}\rightarrow \frac{1}{i}[\hat{x}^{\mu },\hat{x}^{\nu }].
\label{36}
\end{equation}%
The central idea is then quantize the constraint 

\begin{equation}
-\frac{1}{\gamma ^{2}}\sigma ^{\mu \nu }\sigma _{\mu \nu }=T_{p}^{2},
\label{37}
\end{equation}%
which can be derived from (7) setting $p=1$. According to (35) and (36) one
gets

\begin{equation}
([\hat{x}^{\mu },\hat{x}^{\nu }])^{2}=T_{p}^{2}I,  \label{38}
\end{equation}%
where $I$ is the identity operator. It turns out that the constraint (38)
plays an essential role in Matrix theory. Extending the Yoneya's idea for
strings, Oda [14] (see also Ref. [13]) has shown that it is also possible to
construct a Matrix model of M-theory from a Schild-type action for
membranes. It is clear from our previous analysis of identifying the
quantity $\sigma ^{\mu \nu }$ with a chirotope $\chi ^{\mu \nu }$ that these
developments of Matrix theory can be linked with the oriented matroid theory.

Finally, it is known that there are matroids, such as the non-Pappus
matroid, which are not realizable. On the other hand, our discussion on the
present work has been focused in realizable matroid bundles. This suggests
that there must be an extension of $p$-branes of pure combinatorial
character. Moreover, it has been proved that matroid bundles have
well-defined Stiefel-Whitney classes [2] and other characteristic classes
[21]. In turn Stiefel-Whitney classes are closely related to spinning
structures. Perhaps, these exciting developments in combinatorial
characteristic classes may eventually lead to a matroid/supersymmetry
connection.

\bigskip

\noindent \textbf{Acknowledgments: }I would like to thank M. C. Mar\'{\i}n,
J. Saucedo and G. Arreaga for helpful comments.

\bigskip

\smallskip\ \


\begin{thebibliography}{99}
\bibitem{1} R. D. Macpherson, ``Combinatorial differential manifolds: a
symposium in honor of John Milnor's sixtieth birthday,'' pp. 203-221 in
Topological methods on modern mathematics (Stony Brook, NY, 1991), edited by
L. H. Goldberg and A. Phillips, Houston, 1993.

\bibitem{2} L. Anderson and J. F. Davis, ``Mod 2 Cohomolgy of Combinatorial
Grassmannians,'' math.GT/9911158.

\bibitem{3} L. Anderson, New Perspectives. in Geom. Comb. \textbf{38}, 1
(1999).

\bibitem{4} A. Bjorner, M. Las Verganas, N. White and G. M. Ziegler, \textit{%
Oriented Martroids}, (Cambridge University Press, Cambridge, 1993).

\bibitem{5} I. M. Gelfand and R. D. Macpherson, Bull. Amer. Math. Soc. 
\textbf{26}, 304 (1992).

\bibitem{6} J. A. Nieto and M.C. Mar\'{\i}n, Int. J. Mod. Phys. \textbf{A18}%
, 5261 (2003); hep-th/0302193.

\bibitem{7} J. A. Nieto, Rev. Mex. Fis. \textbf{44,} 358 (1998).

\bibitem{8} J. A. Nieto and M. C. Mar\'{\i}n, J. Math. Phys. \textbf{41,}
7997 (2000).

\bibitem{9} J. A. Nieto, J. Math. Phys. \textbf{45}, 285 (2004);
hep-th/0212100.

\bibitem{10} A. Schild, Phys. Rev. \textbf{D 16}, 1722 (1977).

\bibitem{11} J. Gamboa, C. Ram\'{\i}rez and M. Ruiz-Altaba, Nucl. Phys. 
\textbf{B 338},143 (1990).

\bibitem{12} N. Kitsunezaki and S. Uehara, JHEP \textbf{0110}, 033 (2001);
hep-th/0108181.

\bibitem{13} R. Kuriki, S. Ogushi and A. Sugamoto, Mod. Phys. Lett. \textbf{%
A14}, 1123 (1999); hep-th/9811029.

\bibitem{14} I. Oda, "Matrix theory from Schild action"; hep-th/9801085.

\bibitem{15} T. Yoneya, Prog. Theor. Phys. 97, 949 (1997); hep-th/9703078.

\bibitem{16} P. K. Townsend, \textquotedblleft Four lectures on
M-theory,\textquotedblright\ \textit{Proceedings of the ICTP on the Summer
School on High Energy Physics \ and Cosmology}, June 1996, hep-th/9612121.

\bibitem{17} M. J. Duff, Int. J. Mod. Phys. \textbf{A\ 11,} 5623 (1996),
hep-th/9608117.

\bibitem{18} P. Horava and E. Witten, Nucl. Phys. \textbf{B\ 460,} 506
(1996).

\bibitem{19} J. G. Oxley, \textit{Martroid Theory}, (Oxford University
Press, New York, 1992).

\bibitem{20} M. J. Duff, and J. X. Lu, Class. Quant. Grav. \textbf{9}, 1
(1992).

\bibitem{21} D. Biss, \textquotedblleft Some applications of oriented
matroids to topology,\textquotedblright\ PhD. thesis, MIT, 2002.
\end{thebibliography}
\end{document}